\documentclass[12pt]{article}

\usepackage[normalem]{ulem}

\usepackage{xcolor}
\usepackage[english]{babel}
\usepackage[T1]{fontenc}
\usepackage[utf8]{inputenc}
\usepackage{authblk}
\usepackage{mathtools}
\usepackage{slashed}
\usepackage{amsmath,amssymb}
\usepackage{amsfonts}
\usepackage{enumitem}
\usepackage{graphicx,color,xcolor}
\usepackage{cite}
\usepackage{subcaption}
\pdfoutput=1
\usepackage[backref=page]{hyperref}

\hypersetup{
	colorlinks=true,
	linkcolor=blue,
	filecolor=magenta,      
	urlcolor=cyan,
}
\usepackage{cleveref}

\numberwithin{equation}{section} 

\definecolor{refcol}{rgb}{0.9,0.1,0.1}
\hypersetup{colorlinks=true,linkcolor=blue,citecolor=refcol,urlcolor=cyan,linktocpage}

\newcommand{\myfontbackref}[1]{
	\hspace*{\fill} \mbox{\textit {\small #1}}
}

\renewcommand*{\backref}[1]{}
\renewcommand*{\backrefalt}[4]{%
	\ifcase #1 \myfontbackref{no citations}
	\or \myfontbackref{Cited in page #2}
	\else \myfontbackref{Cited #1 times, in pages #2}
	\fi
}

\newcommand{\ben}{\begin{eqnarray}\displaystyle}
\newcommand{\een}{\end{eqnarray}}

\newcommand{\be}{\begin{equation}}
\newcommand{\ee}{\end{equation}}

\newcommand{\bc}{\begin{center}}
\newcommand{\ec}{\end{center}}

\newcommand{\eesp}{\end{split}}
\newcommand{\bsp}{\begin{split}}

\newcommand{\Rmnum}[1]{\expandafter\@slowromancap\romannumeral #1@}
\newcommand{\norm}[1]{|\!| #1 |\!|}

\renewcommand{\d}{\delta}	%%% Redefinition

\newcommand{\cH}{\mathcal{H}}

\newcommand{\cO}{\mathcal{O}}

\newcommand{\cR}{\mathcal{R}}
\newcommand{\cS}{\mathcal{S}}

\newcommand{\expa}[1]{\exp\left( #1 \right)}

\newcommand{\loga}[1]{\log\lb #1 \rb}

\newcommand{\lb}{\left (}
\newcommand{\rb}{\right )}

\newcommand{\bensp}{\begin{eqnarray}\begin{split}}
\newcommand{\eensp}{\end{eqnarray}\end{split}}

\newcommand{\bnm}{\begin{matrix}}
\newcommand{\enm}{\end{matrix}}
\def\XXint#1#2#3{{\setbox0=\hbox{$#1{#2#3}{\int}$ }
\vcenter{\hbox{$#2#3$ }}\kern-.6\wd0}}

\newcommand{\ket}[1]{\big|#1\big>}
\newcommand{\bra}[1]{\big<#1\big|}
\newcommand{\braket}[2]{\big<#1\big|#2\big>}
\newcommand{\expect}[1]{\big<#1\big>}

\newcommand{\dow}{\partial}

\graphicspath{{Plots/}}

\begin{document}
	%%%%%%%%%%%%%%%%%%%%%%%%%%%%%%%%%%%%%%%%%%
	\begin{titlepage}
		\thispagestyle{empty}
		
		\title{
			{\Huge\bf Spread complexity as classical dilaton solutions}
		}
		
		%\bigskip\bigskip\bigskip\bigskip\bigskip
		\vfill
		
		\author{
			{\bf Arghya Chattopadhyay$^a$}\thanks{{\tt \href{mailto:arghya.chattopadhyay@umons.ac.za}{arghya.chattopadhyay@umons.ac.be}}}, {\bf Arpita Mitra$^b$}\thanks{{\tt \href{mailto:arpitamitra89@gmail.com}{arpitamitra89@gmail.com}}},
			{\bf Hendrik J.R. van Zyl$^{c,d}$}\thanks{{\tt \href{mailto:hjrvanzyl@gmail.com}{hjrvanzyl@gmail.com}}}
			\smallskip\hfill\\      	
			\small{
				$^a${\it Service de Physique de l'Univers, Champs et Gravitation, Université de Mons}\\
				{\it 20 Place du Parc, 7000 Mons, Belgium.}\\
				%\hfill
				\hfill\\       
				$^b${\it Department  of  Physics,  Pohang  University  of  Science  and  Technology}\\{\it  Pohang  37673,  Korea.}\\
				%\hfill
				\hfill\\    
				$^c${\it \mbox{Laboratory for Quantum Gravity \& Strings, Department of Mathematics \& Applied Mathematics}}\\{\it University of Cape Town, Cape Town, South Africa.}\\
				\hfill\\
				$^d${\it The National Institute for Theoretical and Computational Sciences }\\{\it Private Bag X1, Matieland, South Africa.}\\
				
			}
		}

		%\bigskip\bigskip\bigskip\bigskip
		\vfill
		
		\date{
			\begin{quote}
				\centerline{{\bf Abstract}}
				{\small We demonstrate a relation between Nielsen's approach towards circuit complexity and Krylov complexity through a particular construction of quantum state space geometry. We start by associating K\"ahler structures on the full projective Hilbert space of low rank algebras. This geometric structure of the states in the Hilbert space ensures that every unitary transformation of the associated algebras leave the metric and the symplectic forms invariant. We further associate a classical matter free Jackiw-Teitelboim (JT) gravity model with these state manifolds and show that the dilaton can be interpreted as the quantum mechanical expectation values of the symmetry generators. On the other hand we identify the dilaton with the spread complexity over a Krylov basis thereby proposing a geometric perspective connecting two different notions of complexity.  }
			\end{quote}
		}
		
		%\vfill
		%\leftline{{\bf Report No: }} }

	%\begin{abstract}
	%\end{abstract}
	
\end{titlepage}
%%%%%%%%%%%%%%%%%%%%%%%%%%%%%%%%%%%%%%%%%%
\thispagestyle{empty}\maketitle\vfill \eject

\tableofcontents
%\newpage

%%%%%%%%%%%%%%%%%%%%%%%%%%%%%%%%%%%%%%%%%%
\section{Introduction}\label{sec:intro}
The notion of \emph{complexity} was introduced initially to measure the efficiency of algorithms for quantum computation by associating different \emph{costs} to different arrangements of quantum gates from a predefined set, needed to go from a particular \emph{reference state} to a \emph{target state} \cite{Nielsen_1,Nielsen_2006,Nielsen_2,Nielsen_3}. Recently, complexity has become an important tool for theoretical physicists to deal with problems ranging from probing the black hole interior \cite{Susskind:2014rva} to diagnosing quantum chaos \cite{Ali:2019zcj,Bhattacharyya:2019txx,Bhattacharyya:2020iic,Bhattacharyya:2020art,Balasubramanian:2021mxo,Balasubramanian:2022tpr} as well as other interesting avenues \cite{Susskind:2014jwa,Caputa:2022eye,Roberts:2014isa,Stanford:2014jda,Susskind:2014moa,Alishahiha:2015rta,Brown:2015bva,Brown:2015lvg,Barbon:2015ria,Hackl:2018ptj,Khan:2018rzm, Mandal:2022ztj}. Due to the freedom of choosing the relevant quantum gates or unitary operators that constitute valid circuits as well as the freedom to choose the reference state and the associated cost, there are several different notions of complexity used in the literature \cite{Yang:2018cgx,Chapman:2018hou,Bhattacharyya:2018wym,DiGiulio:2021oal,Erdmenger:2020sup,Basteiro:2021ene,Parker:2018yvk}. A natural question is whether different measures of complexity may be related in some way and thus capture related physical information.
  In this paper, we will expand upon a geometric relation between Nielsen's approach towards circuit complexity and Krylov complexity  first put forward in \cite{Caputa:2021sib} by demonstrating that the relevant quantities may be obtained as classical solutions of an appropriate action. In particular we have established that the quantum mechanical expectation values of the
symmetry generators in Neilsen's approach can be identified with the spread complexity of a target state in a chosen Krylov basis. \\ \\
Problems in quantum computing can generally be formulated as producing a certain target state from a given initial state by acting on it with some 
 set of accessible quantum gates. Loosely, the notion of \emph{circuit complexity} can be thought of to be the minimum number of gates required to complete this operation. In the continuum limit, the problems in quantum computing gets translated into a geometric problem of finding the minimum path connecting two points in a curved manifold \cite{Nielsen_2006}\footnote{The interested reader may consult \cite{Chapman:2021jbh} and references therein for a detailed review.}. For problems involving quantum field theories, one essentially replaces the set of quantum gates by a set of unitary operators and take Nielsen's geometric approach as mentioned before \cite{Jefferson:2017sdb,Khan:2018rzm,Bhattacharyya:2018bbv}. Since the Nielsen complexity basically computes the minimum number of required gates, in the continuum limit, one has to therefore define a metric in the \emph{space of unitary transformations} and look for an optimised trajectory in this space, minimising the required number of unitary operations, given a pair of reference and target states. At this point one should note that not all possible unitary transformations are allowed in this process otherwise complexity would become trivial. Contrary to its quantum computational counterpart, the Nielsen's approach towards complexity applied to field theory have an inherent ambiguity in the choice of the cost functional associated to a particular path in the manifold of unitary operators. To circumvent this ambiguity in cost function, a subtly different approach can be taken utilising Fubini-Study metric following \cite{Chapman:2017rqy}. Rather than taking the space of unitary transformations, \cite{Chapman:2017rqy} prescribes to take a \emph{space of states} defined by the Fubini-Study(FS) metric. Therefore, in this scenario one has to keep track of the infinitesimal changes in states while preparing the target state from an initial state and the complexity is defined as the geodesic distance between them. More importantly one crucial difference is that while in Nielsen approach every quantum gate or unitary transformation is assigned a fixed cost, FS complexity associates variable costs for each gate depending on the states they act on. For the setups that we will consider there is a one-to-one correspondence between elements of a factor group and quantum states.  As such, the cost function may equivalently be defined on the manifold of unitary transformations or the manifold of target states. \\ \\
In contrast to the considerations above, \cite{Parker:2018yvk} defines a new notion of complexity dubbed Krylov or \emph{spread} complexity, which depends on the \emph{time-evolution} of some reference state in the Hilbert space of the systems under consideration. The time dependence of any operator  is dictated through the Heisenberg equation, which can be formally solved as a power series in time. The basis of these power series expansion are the nested commutators of the operator with the Hamiltonian. The trick is to associate a state corresponding to the operators in the Hilbert space and going over to an orthonormal basis for the power series expansion utilising the Lanczos algorithm. This orthonormal basis is known as the Krylov basis. As elaborated in the following sections, the expansion coefficients in this basis can be thought of as some probability amplitudes whose squared sum is always conserved in time and Krylov complexity is defined to be a weighted sum of these probabilities. Since the operator growth in time as described above, grows exponentially fast at late times, one can further associate analogues of Lyapunov exponents for different quantum systems to measure the bounds on \emph{quantum chaos} \cite{Parker:2018yvk,Murthy:2019fgs}. Recently, Krylov complexity has become the subject of various interesting investigations \cite{Caputa:2021sib,Balasubramanian:2022tpr,Rabinovici:2020ryf,Barbon:2019wsy,Jian:2020qpp,Yin:2020oze,Dymarsky:2019elm,Dymarsky:2021bjq,Kar:2021nbm,vonKeyserlingk:2017dyr,Nahum:2017yvy, Caputa:2022eye,Caputa:2022yju,PhysRevLett.42.1698,Kitaev:2000nmw,Haque:2022ncl,Camargo:2022rnt,Afrasiar:2022efk}. As its other counterparts, Krylov complexity also have an inherent ambiguity while defining the inner products required to establish the Krylov basis, although most of the literature in physics uses the notion defined in \cite{Lanczos1950AnIM}. Nevertheless, the \emph{actual} physical or geometric meaning of the Krylov complexity is still inside a veil.\\ \\
The leitmotif of all notions of complexity thus far, relies on the construction of some geometry. Even before the success of complexity, there are many methods or algorithms to define an underlying geometry of Hilbert space of quantum mechanics. Most notably, using the symplectic structure inherent in quantum mechanics \'a la \cite{Chernoff1974PropertiesOI}, Kibble proposed a \emph{geometrisation of quantum mechanics} in \cite{cmp/1103904831}. Provost and Vallee in \cite{Provost:1980nc} extended this proposal to quantum state space and defined a compatible Riemannian metric on submanifolds of Hilbert space consisting of generalised coherent states\footnote{Consult \cite{Brody:1999cw} and references therein for more details.}. Following Ashtekar and Schilling's viewpoint of K\"ahler structure of quantum field theory \cite{Ashtekar:1997ud}, one of the authors have proposed a dual description of $SU(1,1)$ theories in terms of a two dimensional dilaton gravity theory \cite{Kriel:2015tga}. This otherwise foundational exercise of \emph{geometric reformulation} of quantum mechanics, has interestingly found its application in the literature of complexity starting with \cite{Chapman:2017rqy}. The authors used the same Fubini-Study metric that comes out of the K\"ahler structure of the projective Hilbert space to define complexity. Another hint towards using this quantum state space become apparent through the work of \cite{Caputa:2021sib} where Krylov complexity is shown to be proportional to the volume enclosed by a certain geodesic radius in a hyperbolic disc generated through the coherent states. In this paper we will use the construction described in \cite{Kriel:2015tga} for $SU(1,1)$, $SU(2)$ and Heisenberg-Weyl cases to show that the FS metric described in \cite{Kriel:2015tga} corresponds to the FS complexity and Krylov complexity for $SU(1,1)$ case plays the role of dilaton appearing in the two-dimensional Jackiw-Teitelboim (JT) gravity theory\cite{Jackiw:1982hg,Jackiw:1984je}.\\ \\
This paper is structured as the following. We begin by reviewing a few basics in \cref{sec:basics}, focusing mainly on Nielsen and spread complexity as well as the quantum state manifold which is the setting for our computations. In section {\ref{CaputaResults}} we provide a modest generalisation of the spread complexity computations of \cite{Caputa:2021sib} to include arbitrary choices of reference state, target state and Hamiltonian for unitary representations of low-rank algebras.  Finally, we establish an exact geometric connections between these general spread complexities and Nielsen complexity in \cref{sec:mainresult} through the introduction of a 2d dilaton model. We conclude in \cref{sec:conclusion}, with possible generalities and future directions. We also include the basic origin of metric and the symplectic structure of quantum state manifolds in \cref{app:qstate}, along with a simple example. In \cref{UnnormalisableRefStates}, we discuss manifolds for non-normalisable states and their isometries and Killing vectors.

\section{A few basics}\label{sec:basics}
In the following two subsections we will briefly review the two core ingredients of our proposal as mentioned before. Apart from increasing the pedagogic value for the benefit of readers, this would also serve the purpose of setting up our notations and conventions.

\subsection{Complexity}
Our focus will be on continuous notions of complexity for quantum states (as opposed to operator complexity).   In several formulas it will turn out to be useful to drop normalisation factors.  We will use the notation $|\cdot)$ and $(\cdot|\cdot)$ to denote unnormalised (and potentially non-normalisable) vectors and their inner product and reserve the notation $|\cdot\rangle$ for normalised vectors.  

\subsubsection{Nielsen complexity}

The Nielsen approach \cite{Nielsen_1,Nielsen_2, Nielsen_2006,Nielsen_3} considers the set of all possible unitary gates to be elements of a unitary representation of some symmetry group, $U \in G$.  These may act on a reference state $|\phi_r\rangle$ to give the Hilbert space of accessible target states
\begin{equation}
|\phi_t\rangle = U(\sigma) |\phi_r\rangle.    \label{targetState}
\end{equation}
In the above we have introduced a circuit parameter $\sigma$, usually taken to run from $0$ to $1$ so that $U(0) = I$.  Written in terms of the generators of the group, $A_i$, the target state is given by 
\begin{equation}
|\phi_t(s_1, s_2, \cdots, s_n)\rangle   = e^{i \sum_j s_j(\sigma) A_j} |\phi_r\rangle. 
\end{equation}
Note that various choices of the functions $s_j(\sigma)$ can be made to give circuits that connect the reference state and desired target state.  When computing complexity one will make a choice for these functions that minimise some choice of cost function.  The cost function can be implemented as a metric defined on the Hilbert space of target space (or equivalently on the space of unitary gates).  This choice is not unique, but a popular choice is the Fubini-Study metric
\begin{equation}
ds^2 = \sum_{j,k} \left( \langle \phi_r | \frac{d U^\dag }{d s_k} \frac{d U }{d s_j} |\phi_r\rangle - \langle \phi_r | \frac{d U^\dag }{d s_k} U |\phi_r\rangle  \langle \phi_r | U^\dag \frac{d U }{d s_j} |\phi_r\rangle   \right) \frac{d s_k}{d\sigma} \frac{d s_j}{d \sigma} d\sigma^2. \label{FSdef}
\end{equation}
An appealing feature of this cost function is that, assuming that
all symmetry transformations are equally easy to perform, the cost function for the class
of simple gates we consider is fixed up to a global choice of units \cite{Magan:2018nmu}.  The choices of the functions $s_j(\sigma)$ that minimise the cost function correspond to geodesics of the metric 
\begin{equation}
ds^2 = \sum_{j,k} \left( \langle \phi_r | \frac{d U^\dag }{d s_k} \frac{d U }{d s_j} |\phi_r\rangle - \langle \phi_r | \frac{d U^\dag }{d s_k} U |\phi_r\rangle  \langle \phi_r | U^\dag \frac{d U }{d s_j} |\phi_r\rangle   \right) d s_k d s_j    \label{FSMet2} 
\end{equation}
and the complexity gets mapped to the geodesic distance.  \\ \\ 
Note that group elements from the stationary subgroup, $H$, of the reference state $|\phi_r\rangle$ i.e. those that act as 
\begin{equation}
    U|\phi_r\rangle = e^{i \theta} |\phi_r\rangle
\end{equation}
add zero cost through (\ref{FSMet2}).  As such,  states that differ by an overall phase should be identified with the same point on the manifold of accessible target states.  At this point is worthwhile to note that the space of accessible target states, as defined through (\ref{targetState}), are precisely generalized coherent states as introduced by Perelemov \cite{Perelomov}.  There is a one-to-one correspondence between accessible target states and elements of the factor group $G/H$.  We elaborate on this point in subsection \ref{sec:geoqu}.

\subsubsection{Spread complexity}

Another method by which to quantify the computational cost associated with preparing a desired target state $|\phi_t\rangle$ is spread complexity \cite{Balasubramanian:2022tpr}.  The central idea here is that there exists an \emph{ordered} basis (ordered in increasing complexity) for the Hilbert space.  In terms of such a basis $|B_n\rangle$ the spread complexity of a desired target state is given by
\begin{equation}
C = \sum_n c_n |\langle \phi_t | B_n\rangle|^2,    \label{spreadCost}
\end{equation} 
where the weights $c_n$ are strictly increasing with $n$.  \\ 
A physically motivated algorithm to obtain such a basis is as follows \cite{Caputa:2021sib}:  Consider the time-evolved reference state 
\begin{equation}
|t \rangle = e^{i t H}|\phi_r\rangle,
\end{equation}
with some Hamiltonian of interest, $H$.  The accessible target states are taken to be any linear combination of time-evolved reference states.  A natural (unnormalised) basis one may then write down for the Hilbert space of target states is 
\begin{equation}
|\mathcal{O}_n ) = \frac{1}{n!}(i H)^n |\phi_r\rangle.
\end{equation}
These states are not, however, orthogonal with respect to the Hilbert space inner product.  They can be orthogonalised by means of a Gram-Schmidt process.  A way to implement this procedure iteratively for Hermitian operators is the Lanczos algorithm \cite{Lanczos1950AnIM} which yields the orthonormal Krylov basis $|K_n\rangle$ \cite{Parker:2018yvk}
\begin{eqnarray}
    |K_0\rangle & \equiv & |\phi_r\rangle\,, \nonumber \\
    |A_{n+1}) & = & (H - a_n) |K_n\rangle - b_n |K_{n-1}\rangle\,, \nonumber \\
    |K_n\rangle & \equiv &  (A_n| A_n)^{-\frac{1}{2}}  |A_n),   \label{Krybasis}
\end{eqnarray}
with the Lanczos coefficients defined as follows,
\begin{eqnarray}
a_n & = & \langle K_n | H |K_n \rangle\,, \nonumber \\
b_n & = & (A_n| A_n)^{ \frac{1}{2}   }\,.
\end{eqnarray}
The Hamiltonian is tri-diagonal in the Krylov basis,
\begin{equation}
    H |K_n\rangle = a_n |K_n\rangle + b_n |K_{n-1}\rangle + b_{n+1} |K_{n+1}\rangle\,,
\end{equation}
and it has been demonstrated that the Krylov basis minimises the complexity of the time-evolved reference state for cost functions of the form (\ref{spreadCost}) \cite{Balasubramanian:2022tpr}.  \\ \\
With the basis fixed $|B_n\rangle = |K_n\rangle$ one now needs to fix the coefficients $c_n$ in the cost function (\ref{spreadCost}).   A popular choice is simply $c_n =n$ so that the spread complexity of a target state $|\phi_t \rangle$ in the Krylov basis is defined as
\begin{equation}
    K = \sum_{n=0}^\infty n  \langle \phi_t | K_n \rangle  \langle K_n| \phi_t \rangle\,,\label{SC}
\end{equation}
which is the expectation value of the Krylov complexity operator with respect to the target state
\begin{equation}
\hat{K} = \sum_{n=0}^\infty n |K_n\rangle \langle K_n |.   \label{KCompOperator}
\end{equation}
The Lanczos algorithm allows one (at least in principle) to compute spread complexity for an arbitrary choice of Hamiltonian and reference state.  In this paper we will be interested in the case where the Hamiltonian is the element of some symmetry algebra.  
\\ \\
In these special cases, the Krylov subspace can be represented as a one-dimensional subspace on the manifold of target states of the Nielsen complexity for the same symmetry group.  Indeed, the Krylov subspace generated by the Hamiltonian on any choice of reference state on the manifold may be represented in a similar way.  It thus follows that the manifold of accessible target states associated with the symmetry group is a natural setting for both notions of complexity. This brings in the necessary perspective to briefly review the \emph{geometric reformulation} of quantum mechanics in the next subsection.

\subsection{Geometry of quantum state manifolds}\label{sec:geoqu}

We will now make some general statements regarding quantum state manifolds, for the benefit of the interested reader. Unlike its counterpart, classical mechanics is intrinsically geometric in nature. One can visualise classical states being points on the phase space $\Gamma$, which is also a symplectic manifold. The classical observables are then described as some real-valued function on this manifold. Being symplectic, the \emph{classical state manifold} already have a Lie-bracket, which is the Poisson bracket. Each individual observable $f$ can also be associated with a Hamiltonian vector field, generating individual flows on $\Gamma$. The flow corresponding to the Hamiltonian $H$ describes the time evolution of the system. \\ \\
 On the flip side, the \emph{canonical} language of quantum mechanics is rather algebraic, revolving around the construction of a Hilbert space $\cH$. Each state is described by a ray in the Hilbert space, and observables are the self-adjoint linear operators defined on $\cH$ \cite{Ashtekar:1997ud}. The Lie-algebra appears through the commutators between different observables. Similar to classical mechanics, operators here also generates a flow on the state-space and the dynamics is described through the flow corresponding to the Hamiltonian operator $\hat{H}$. Although one should note that, in this case the flow is generated through a one parameter group $\exp(i\lambda \hat{G})$ corresponding to some observable $\hat{G}$. \\ \\
Therefore, it is quite interesting to look for geometric structures of quantum mechanics as well. The main interest in this approach kindled with Kibble's work in \cite{cmp/1103904831}, where he showed that Schr\"odinger equation can be regarded as a Hamiltonian flow on the Hilbert space $\cH$. Heslot in \cite{heslot} observed that the Hilbert space $\cH$ admits a symplectic formulation where the projective Hilbert space behaves analogous to the phase space. There are several works in this direction mostly restricted to the scenario of finite dimensional Hilbert spaces \cite{Gibbons:1991sa,Hughston,Hughston1996GeometryOS}. In the current situation we will consider the geometric structure of $\cH$ for infinite dimensional cases \cite{Renzo1,Renzo2,Ashtekar:1997ud,Kriel:2015tga}. In the following, we will follow \cite{Kriel:2015tga} to set up our notations and conventions\footnote{See \cite{vanZyl:2015dsa} for further details and references on this topic.} for the sections to come. The origin of the metric and the symplectic form of the projective Hilbert space is reviewed in \cref{app:qstate}.

\subsubsection{Metric and the symplectic structure}
We will start by considering a Hilbert space $\cH$ with a family of normalised state vectors $\cS\equiv\{\ket{s}\}$ such that $\cS\subseteq\cH$ and the state vectors are smoothly parametrised by a set of coordinates $s=\left(s_1,s_2,\cdots,s_n\right)\in \mathbb{R}^n$. For these vectors to have \emph{physical significance}, they have to be invariant under the transformation $\ket{s}\rightarrow e^{i \phi(s)}\ket{s}$, with $\phi(s)$ being some real function of $s$. Therefore, one can associate with each $\ket{s}$, a ray $\widetilde{\ket{s}}=\{e^{i\theta}\ket{s}:\theta\in\mathbb{R}\}$, and all elements of a ray represents the same physical content. As shown in \cref{app:qstate}, given the definition of inner product in this context, the metric $g_{ij}$ and the two form $\sigma_{ij}$ can be written as 
\begin{equation}\label{eq:sympdef}
	g_{ij}(s)=\left.\dow_i\dow_j'\loga{\braket{s}{s'}}\right|_{s=s'},\quad \text{and}\quad \sigma_{ij}={1\over 2i}\left.\left(\dow_i\dow_j'\log{\braket{s}{s'}\over \braket{s'}{s}}\right)\right|_{s=s'},
\end{equation} 
with $\dow_i\equiv{\dow\over \dow s_i}$. This expression of the metric and the 2-form is also invariant under any arbitrary scaling of the states by some scalar function. This invariance is very crucial in regulating the inner product for manifolds generated through potentially non-normalisable states. This construction is briefly reviewed in \cref{UnnormalisableRefStates}.
\\ \\
Interestingly, for manifolds of (potentially) non-normalised states $\cS$, parametrised holomorphically by a set of complex coordinates $z=\left(z_1,z_2,\cdots,z_n\right)\in \mathbb{C}^n$, the non-zero components of $g$ and $\sigma$ can be written as 
\begin{equation}\label{eq:cmplxdef}
	g_{a\bar{b}}=g_{\bar{b}a}={1\over 2}\dow_a\dow_{\bar{b}}\log(\bar{z}|z)\quad \text{and}\quad \sigma_{a\bar{b}}=-\sigma_{\bar{b}a}=ig_{a\bar{b}}.
\end{equation}
with $\dow_a\equiv{\dow\over \dow z_a},\,\dow_{\bar{a}}\equiv{\dow\over \dow_{\bar{z}_a}}$ with $a,b\in[1,n]$. Since the 2-form is non-degenerate, it also qualifies as the symplectic form and thus $\cS$ is a K\"ahler manifold with potential ${1\over 2}\log(z|z)$. Further, assuming $z_j=a_j+ib_j$, the tangent space at $|z)$ can be identified with $\text{span}_{\mathbb{R}}\{\dow_{a_i}|z),\dow_{b_i}|z)\}$ which remains invariant under multiplication of $i$ since $i\dow_{a_i}|z)=\dow_{b_i}|z)$. Thinking of this $n$ dimensional complex manifold as $2n$ dimensional real manifold paramaterised by $s=\{s_1,s_2,\cdots,s_{2n}\}$ this same relation can be expressed through the complex structure $J$ as
$$i\dow_{k}|s)={J^j}_k\dow_j|s).$$
The complex structure $J$ has the property that ${J^i}_k{J^k}_j=-\delta^i_j$ so that $J^2=-1$. Following \cite{Kriel:2015tga}, on can further show that
\begin{equation}\label{eq:cmplxj}
	\sigma_{ij}={J^k}_ig_{kj}\quad \text{and} \quad J_{ij}=\sigma_{ji}.
\end{equation}
Using \eqref{eq:cmplxdef} it can then be written
\begin{equation}
	{J_a}^{\bar{b}}={J_a}^{\bar{b}}=0 \quad \text{and}\quad {J_a}^b=-{J_{\bar{a}}}^{\bar{b}}=i\delta_{a}^b.
\end{equation}
In the following we will focus on the manifolds generated through the action of a Lie group on a fixed reference state as discussed in the \cref{app:qstate}. 
\subsubsection{Vector field action}
Let us now consider an operator $\hat{G}$ on the state manifold, which generates a transformation that leave the manifolds of rays $\widetilde{\cS}=\{\widetilde{|s\big>}\}$ invariant i.e. 
\begin{equation}
e^{i \lambda \hat{G}} |s\rangle = e^{i \phi(s)} |u(s) \rangle.
\end{equation}
The infinitesimal version of this transformation implies the existence of a scalar function $\phi(s)$ and vector field $X_{\hat{G}}=k^i\dow_i$ such that
\begin{equation}\label{eq:sym}
	\hat{G}\ket{s}=\phi(s)\ket{s}-iX_{\hat{G}}\ket{s}
\end{equation}
using \eqref{eq:cmplxj}, then one can show that \cite{Kriel:2015tga},
\begin{equation}
	k^i=-{1\over 2}\left[\sigma^{ij}\dow_j\expect{\hat{G_1}}+g^{ij}\dow_j\expect{\hat{G_2}}\right],  \label{vecField}
\end{equation}
with $\hat{G}=\hat{G_1}+i\hat{G_2}$ and $\expect{\cdot}=\bra{s}\cdot\ket{s}$. Which further simplifies in the complex coordinates as 
\begin{equation}
	k^a=-{1\over 2}\sigma^{a\bar{b}}\dow_{\bar{b}}\expect{\hat{G}},\quad \text{and}\quad k^{\bar{a}}=-{1\over 2}\sigma^{\bar{a}b}\dow_{b}\overline{\expect{\hat{G}}}  
  \label{KillingExpectationVal}.
\end{equation}
Therefore the vector field includes both the information of the Riemannian and the symplectic structure of the manifold. Evidently, one can consider $\expect{\hat{G}}$ as a scalar field on the manifold. In fact, \eqref{eq:sym} further implies that 
\begin{equation}
	X_{\hat{G}}\expect{\hat{\cO}}=i\expect{\hat{\cO}\delta\hat{G}-\delta\hat{G}^\dagger\hat{\cO}},
\end{equation}
where $\hat{\cO}$ is some arbitrary operator on the manifold and $\delta\hat{G}=\hat{G}-\expect{\hat{G}}$. In case, $\hat{G}$ is Hermitian then,
\begin{equation}\label{eq:killingvector}
	X_{\hat{G}}\expect{\hat{\cO}}=\expect{[-i\hat{G},\hat{\cO}]}\quad \text{and}\quad X_{i\hat{G}}\expect{\hat{\cO}}=\expect{\{    -\delta\hat{G},\hat{\cO}  \}  }.
\end{equation}
Hence, one can infer that if the symmetries of $\tilde{\cS}$ is generated by a Lie algebra with generators being the set of Hermitian operators $\{\hat{D}_i\}$, then the vector fields $\{X_{\hat{D}_i}\}$ provides a representation of the Lie algebra spanned by $\{-i\hat{D}_i\}$ and the scalar fields $\{\expect{\hat{D_i}}\}$ transforms under the adjoint representation. \\ \\
In summary, the manifold of accessible target states, equipped with the Fubini-Study metric, allows for the action of operators to be represented by vectors fields.  In particular, the symmetry generators are associated with  Killing vector fields on the manifold.  We will make profitable use of this observation later.

\section{Complexity for low-rank algebras}

\label{CaputaResults}

In \cite{Caputa:2021sib} an elegant interpretation of Krylov complexity was found as a volume contained  on the information metric for $SL(2,R), SU(2)$ and Heisenberg-Weyl symmetry groups.  Additionally, the authors demonstrated that, for these examples, the Krylov complexity operator (\ref{KCompOperator}) can be identified with an isometry of the information metric.  \\ \\
For our later use, we provide here a modest generalisation of these results.  We consider a low-rank algebra spanned by one pair of ladder operators and their commutator, $L_{+}, L_{-}, \left[L_{-}, L_{+} \right]${\footnote{These operators are denoted as $K$ in section \ref{kill} for the $su(1,1)$ algebra.}}.   The algebra we consider is
\begin{eqnarray}
L_{+} & = & L_{-}^\dag, \nonumber \\
\left[ \left[L_{-}, L_{+} \right], L_{+} \right] & = & 2 f L_{+},   \label{algebra} \\ 
\left[ \left[L_{-}, L_{+} \right], L_{-} \right] & = & -2 f L_{-},    \nonumber
\end{eqnarray}
parametrised by the real structure constant $f$. To recover $su(1,1), su(2)$ and the Heisenberg-Weyl algebra we have $f=1,-1$ and $0$ respectively. \\ \\
To proceed with the computation  (both of state spread complexity and Nielsen complexity) we need to specify a reference state.  A possible choice is the highest weight state defined by
\begin{eqnarray}
L_{-} |w\rangle & = & 0,   \nonumber \\
\left[ L_{-}, L_{+} \right] |w\rangle & = & w_0 |w\rangle,\nonumber
\end{eqnarray}
which we will specify as the reference state $|\psi_0 \rangle = |w\rangle$.  The reason for the subscript $0$ will be clear shortly.  Any group element acting on this reference state may be parameterised as 
\begin{equation}
e^{i (  a_{+} L_{+} + a_{+}^{*}L_{-} + a_0 [L_{-}, L_{+}]  )} |\psi_0\rangle = N e^{z L_{+}}|\psi_0\rangle,
\end{equation}
where $z$ is a complex coordinate that depends on the parameters $a_{+}$ and $a_{0}$. As highlighted in section \ref{sec:basics}, this is precisely a generalized coherent state \cite{Perelomov} and these states are in a one-to-one correspondence with elements of $G/([L_{-}, L_{+}] )$.  This manifold of states represents the set of accessible target states.  The reference state is also represented by a point on this manifold, in this case $z=0$.   \\ \\
We compute the Fubini-Study metric for these low-rank algebras  following \eqref{FSdef} as 
\begin{eqnarray}
ds^2 & = &  \partial_z \partial_{\bar{z}}  \log\left(  \langle w | e^{\bar{z} L_{-}} e^{z L_{+}} |w\rangle     \right) dz d\bar{z}    \nonumber \\
& = & \partial_z \partial_{\bar{z}}\log \left( 1 - f z \bar{z}  \right)^{-\frac{ w_0}{f}}  dz d\bar{z} \nonumber \\
& = & \frac{ w_0 }{(1 - f z \bar{z})^2} d z d\bar{z}.
\end{eqnarray}
  This is a metric of constant scalar curvature given by
\begin{equation}
R = -\frac{8f}{w_0}.    \label{FScurvature}
\end{equation}
Note that the scalar curvature is dependent on the algebra structure constant and the weight of the reference state.  In particular (for positive $w_0$) we recover Euclidean $AdS_2$, $dS_2$ and flat geometries for the $su(1,1)$, $su(2)$ and Heisenberg algebras respectively, in agreement with \cite{Caputa:2021sib}. One can easily check that a parametrization
\begin{equation}
   z(\sigma)=re^{i\theta}, 
\end{equation}
with $w_0=4$ and $f=1$ gives us the JT metric with disk topology,
\begin{equation}
     ds^2=\frac{4}{(1-r^2)^2}(dr^2+r^2 d\theta^2).
\end{equation}
In this computation we need not have selected the highest weight state as our reference state.  Instead, one may have taken the reference state to be 
\begin{equation}
|\psi_{z_0}\rangle = U(\bar{z}_0, z_0) |w\rangle,   \label{generalRefState}
\end{equation}
where $U(\bar{z}_0, z_0) $ is a representative of $G/([L_{-}, L_{+}])$ and may be parametrised as
\begin{equation}
U(\bar{z}_0, z_0) = e^{z_0 L_{+}} e^{\frac{1}{2 f} \log(1 - f \bar{z_0} z_0)[L_-,L_+]} e^{-\bar{z}_0 L_{-}}.
\end{equation}
When we set $z_0 =0$, the reference state (\ref{generalRefState}) is simply the highest weight state. On the manifold of states, the action of the unitary $U(\bar{z}_0, z_0)$ has the effect of shifting the point on the manifold of states that corresponds to the reference state.   For Nielsen complexity this transformation has a rather trivial effect of simply  reorganising the points on the manifold by means of a coordinate transformation. The geodesic distance between points is insensitive to the choice of coordinates, so that Nielsen complexity between reference and target states is unaffected.  \\ \\
The situation is different for spread complexity, however.  This is because the Krylov basis (and therefore the measure of complexity itself) is altered by selecting a different reference state. To compute spread complexity we first need to determine the Krylov basis. This requires a choice of Hamiltonian which we may take as a general algebra element
\begin{equation}
\alpha L_{+} + \alpha^* L_{-} + \gamma [L_{-}, L_{+}],   \nonumber
\end{equation}
and reference state, which we take to be the general $|\psi_{z_0}\rangle$ in (\ref{generalRefState}).  
Regardless of the choice of coefficients\footnote{This is special for the low-rank algebras we consider in this article.  In general, the Krylov basis will depend on the coefficients appearing in the Hamiltonian, see e.g. \cite{Haque:2022ncl}.}, we obtain the Krylov basis and Krylov complexity operator from \eqref{Krybasis} and \eqref{KCompOperator} as
\begin{eqnarray}
|K_{n}\rangle & = & \frac{U(\bar{z}_0, z_0) (L_{+})^n |w\rangle}{     \sqrt{ \langle w | (L_{-})^n (L_{+})^n |w\rangle   }},   \nonumber \\
\hat{K} & = & \sum_{n} n \frac{ U(\bar{z}_0, z_0) (L_{+})^n |w\rangle \langle w | (L_{-})^n U^\dag(\bar{z}_0, z_0)     }{\langle w | (L_{-})^n (L_{+})^n |w\rangle }.
\end{eqnarray}
To compute the spread complexity of an arbitrary target coherent state (built from the highest weight state) we may write
\begin{equation}
|z) = e^{z L_{+}} |w\rangle = U(\bar{z}_0, z_0) e^{z' L_{+}} |w\rangle,
\end{equation}
at the cost of a coordinate transformation
\begin{equation}
z' = \frac{z - z_0}{1 - f z \bar{z}_0}.   \label{coordTrans}
\end{equation}
Note the use of the round bracket since $|z)$ is not normalised.  The normalised states are given by
\begin{equation}
|z\rangle = (\bar{z}|z)^{-\frac{1}{2}} |z).
\end{equation}
We can now readily compute its spread complexity by taking the expectation value of the Krylov complexity operator w.r.t. this state and find
\begin{eqnarray}
\frac{(\bar{z} | \hat{K} | z   )}{(\bar{z} | z)} & = & \sum_{n} n \frac{(\bar{z}' z')^n}{(n!)^2} \frac{ \langle w| (L_{-})^n (L_{+})^n |w\rangle \langle w | (L_{-})^n (L_{+})^n |w\rangle     }{ (\bar{z}' | z') \langle w | (L_{-})^n (L_{+})^n |w\rangle }   \nonumber \\
& = & \sum_{n} n \frac{(\bar{z}' z')^n}{(n!)^2} \frac{ \langle w|(L_{-})^n (L_{+})^n |w\rangle    }{ (\bar{z}' | z')  }   \nonumber \\
& = & z' \partial_{z'} \log( (\bar{z}' | z')    ).
\end{eqnarray}
This may be expressed, in general, as the expectation value of some algebra element as follows.  We note that
\begin{equation}
[ L_{-}, L_{+}   ] e^{z' L_{+}} |w\rangle = (w_0 + 2f z' \partial_{z'} ) e^{z' L_{+}} |w\rangle,
\end{equation}
so that the spread complexity (in the Krylov basis) of the coherent state $|z)$ is given by
\begin{eqnarray}
\frac{(\bar{z} | \hat{K} | z   )}{(\bar{z} | z)}  & = & \langle \bar{z}' | \frac{  1   }{2 f} (\left[ L_{-}, L_{+}  \right] -   w_0 ) | z' \rangle \nonumber \\
& = & \langle \bar{z} | \frac{  1   }{2 f} (  U(\bar{z}_0, z_0) \left[ L_{-}, L_{+}  \right] U^\dag(\bar{z}_0, z_0) - w_0 ) | z \rangle.
\end{eqnarray}
Note that the unitary transformation performed on the reference state now features in the expectation value.  Depending on the unitary transformation relating the reference state for Nielsen complexity and reference state for Krylov complexity, we need to compute the expectation value of a different operator with respect to the target state.  \\ \\
One may proceed with the computation in several ways but ultimately we find that coherent state $|z\rangle$ with the Krylov basis built on the reference state $U(\bar{z}_0, z_0) |w\rangle$ is given by
\begin{eqnarray}
K(|z\rangle; H, U(\bar{z}_0, z_0)|w\rangle ) & = & z' \partial_{z'} \log (\bar{z}' | z') \nonumber \\
& = &  \frac{w_0 z' \bar{z}'}{1 - f z' \bar{z}'} \nonumber \\
&=&\frac{ w_0 (z - z_0)(\bar{z} - \bar{z}_0)  }{ (1 - f \bar{z}_0 z_0)(1 - f \bar{z} z)}.   \label{spreadGen}
 \end{eqnarray}
In the above we have emphasised that the Krylov basis (and thus the associated spread complexity of states) is a function of both the choice of Hamiltonian\footnote{Though, as we have seen, for this simple group the specific choice of Hamiltonian from the algebra plays almost no role, since the Krylov basis is independent of its coefficients.} and the choice of reference state.  The reference state is, by definition, set to zero complexity.  This is reflected nicely in the above formula where the spread complexity vanishes for $z = z_0$.  Note that we recover the expected formulas (such as those in \cite{Caputa:2021sib}) when we set $z_0 = 0$.  We will soon see in the next section, how this formula arises from the matter-free JT gravity model.  \\ \\
Before this we would like to comment how the interpretation of \cite{Caputa:2021sib} (of spread complexity is geodesic volume) holds true for these general Krylov subspaces.  As mentioned, the Krylov subspace can always be represented as a one-dimensional submanifold of the manifold of quantum states.  Consider, for example, a general reference state (coherent state) time-evolved subjected to the Hamiltonian $H = \alpha(L_{+} + L_{-})$.  By making use of Baker-Campbell-Haussdorff equations we can represent the time-evolved state as a $t$-dependent family of coherent states
\begin{equation}
|t\rangle = e^{i \alpha t (L_{+} + L_{-}) } U(\bar{z}_0, z_0) |w\rangle =N e^{ \left( z_0 + i \frac{(1 + f z_0^2) \sinh( \sqrt{f} \alpha t  )   }{\cosh(\sqrt{f} \alpha t ) - i f z_0 \sinh(\sqrt{f} \alpha t)  }    \right) L_{+}   } |w\rangle.  
\end{equation}
Plugging in the value of $z$ of the above into the expression for the spread complexity (\ref{spreadGen}) yields
\begin{equation}
K(|t\rangle; H, U(\bar{z}_0, z_0)|w\rangle ) = \frac{w_0 (1 + f z_0^2)(1 + f \bar{z}_0^2) }{f (1 - f \bar{z}_0 z_0)^2  } \sinh^2(\sqrt{f} \alpha t)
\end{equation}
When comparing this to the geodesic volumes of \cite{Caputa:2021sib} we note, interestingly, that this quantity is also proportional to it.  \\ \\
The appropriate family of states to compare, however, are realised after a unitary transformation i.e. we consider the time evolution of the reference state
\begin{equation}\label{Uprime}
U|\psi_0\rangle 
\end{equation}
under the Hamiltonian
\begin{equation}
H' = U ( \alpha L_{+} + \alpha L_{-}     ) U^\dag   \label{Hprime}
\end{equation}
The family of coherent states involved in this evolution is parametrised as
\begin{equation}
e^{i \alpha t H'} U(\bar{z}_0, z_0)|w\rangle = e^{\frac{- i \sqrt{f} z_0 \cosh(\alpha\sqrt{f} t) + \sinh(\alpha \sqrt{f} t)}{-i \sqrt{f} \cosh(\alpha \sqrt{f} t) + f z_0 \sinh(\alpha \sqrt{f} t)} L_{+}  } |w\rangle    \label{CohStateTime}
\end{equation}
which gives the spread complexity as 
\begin{equation}
K( e^{i t H'} U |w\rangle, H, U |w\rangle   ) = \frac{w_0}{f} \sinh^2(\alpha \sqrt{f} t)
\end{equation}
The appropriate Fubini-Study metric to use is also built on the unitarily transformed generators i.e.
\begin{equation}
U'(z, \bar{z}) = U(\bar{z}_0, z_0) U(\bar{z}, z) U^\dag(\bar{z}_0, z_0)
\end{equation}
from where we may compute the generating function for the Fubini-Study metric as
\begin{equation}
\left( \langle w| U^\dag(\bar{z}_0, z_0)  
 \right)U'^\dag(\bar{z}', z') U'(\bar{z}, z) \left( 
 U(\bar{z}_0, z_0) |w\rangle   \right)
\end{equation}
This (rather trivially) gives rise to the same FS metric in the $z'$ coordinates, but note that the origin of the geometry now represents the state $|z_0\rangle$ and not $|w\rangle$.  The coherent state coordinate $z$ is related to $z'$ as (\ref{coordTrans}).  The time-evolved states (\ref{CohStateTime}) correspond to the coordinates 
\begin{equation}
z' = e^{i \frac{\pi}{2} } \frac{\tanh(\sqrt{f} \alpha t)}{\sqrt{f}}
\end{equation}
and corresponds to geodesic motion.  Indeed, the expressions one obtains in the $z'$ coordinates matches precisely with those of \cite{Caputa:2021sib}.  The geodesic volume computed from the FS metric in the $z'$ coordinate is the appropriate quantity to match with reference state (\ref{Uprime}) time-evolved under the Hamiltonian (\ref{Hprime}).  \\ \\
This demonstrates that the connection between geodesic volume and spread complexity of \cite{Caputa:2021sib} generalises to arbitrary choices of coherent state as reference states.  To be specific, one has to match up the reference state of spread complexity with the center of the Fubini-Study metric and perform the corresponding unitary transformation on the Hamiltonian.  In what follows we will not make these specific choices, but instead discuss the spread complexity for arbitrary choices of reference state, target state and Hamiltonian.

\section{Complexity from Dilaton gravity}\label{sec:mainresult}

We now demonstrate how, using a duality construction first utilised in \cite{Kriel:2015tga}, we can recover both the Fubini-Study metric characterising the Nielsen complexity and spread complexity as classical solutions of a 2d dilaton gravity model.   The relevant action is of the form
\begin{equation}
	S={1\over 2\pi}\int d^2x\,\sqrt{g}\left[\eta \cR+V(\eta)\right] + S_{\text{boundary}},
\end{equation}
with $\eta$ being the dilaton and $V(\eta)$ being some arbitrary potential of the dilaton. The (bulk) equations of motion are 
\begin{equation}
	\cR=-V'(\eta)\quad \text{and}\quad \nabla_\mu\nabla_\nu\eta-{1\over 2}g_{\mu\nu}V(\eta)=0.\label{dileom}
\end{equation}
A generic \emph{static} solution for $g$ and $\eta$ in this case \cite{Kriel:2015tga} can be written as\footnote{This equation matches exactly with \eqref{eq:metric2d} if one maps $(t,\beta)\rightarrow(t,r)$, where 
$r=-{1\over 2}F'(\beta)=\bra{\beta}\hat{H}\ket{\beta}.$  Therefore we can also see that the dilaton $\eta$ coincides with the expectation value of the Hamiltonian which is also the isometry generator for the static case.}
\begin{equation}
	g=\bar{C}(r)dt^2+{dr^2\over 4 \bar{C}(r)};\quad \eta=r;\quad V(r)=4\bar{C}'(r).
\end{equation}
The manifolds of coherent states considered have a high degree of symmetry which restricts the scalar curvature to be a constant.  As such we specialise to a model with a linear potential $V = -R \eta$.  This is the famous Jackiw-Teitelboim model \cite{Jackiw:1982hg,Jackiw:1984je}. The equation of motion for $\eta$ pertaining to more general (non-static) solution of the gravity theory in \eqref{dileom} can be recast as 
\begin{equation}
	\nabla^2\eta+\cR\eta=0\quad \text{and}\quad \nabla_\mu k_{\nu}+\nabla_\nu k_\mu=0
\end{equation}
where $k^{\mu}=-{1\over 2}\sigma^{\mu\nu}\nabla_\nu\eta$. This establishes a relation between the dilaton and the Killing vectors.  This relation along with equation (\ref{KillingExpectationVal}) demonstrates that the dilaton may be matched with the expectation values of symmetry generators!  This relation has already been pointed out in \cite{Kriel:2015tga}  and a detailed discussion is given in Appendix \ref{kill}. In the present context, we make the underlying connection with quantum computational complexity explicit.   \\ \\
As an interesting aside comment for later, we take note of the following conserved quantity
\begin{equation}
M = -\frac{1}{2}\left( (\nabla \eta)^2 + \frac{R}{2}\eta^2  \right)    \label{JTEnergy}
\end{equation}
which is traditionally identified with the mass or energy of the gravitational system \cite{Mann:1992yv, Navarro:1997gr}. 
\\ \\
To realize the spread complexity for low-rank algebras as the classical dilaton solution of a matter-free JT gravity theory we have to impose appropriate boundary conditions. The specific action we need to consider is given by
\begin{equation}
	S={1\over 2\pi}\int d^2x\,\sqrt{g}\eta \left(R + \frac{8f}{w_0} \right)
\end{equation}
The equation of motion resulting from varying the action w.r.t. the dilaton fixes the scalar curvature to be a constant.  In two dimensions, the classical background geometry is thus fixed up to a coordinate transformation to correspond to Fubini-Study metric with curvature (\ref{FScurvature}).This background geometry can be matched with the corresponding manifold of accessible target states.  As such the geodesic distances on this background thus captures the Nielsen complexity between points representing reference and target states.
 \\ \\
 Focusing now on the equations that follow from varying the action w.r.t. the metric, the dilaton (\ref{dileom}) permits the general solution
\begin{equation}
\eta(z, \bar{z}) = \frac{c_1 z + c_2 \bar{z}  + c_3 ( f z \bar{z} + 1)}{1 - f z \bar{z}}\label{classsoln}
\end{equation}
The reality of dilaton would further pose the restriction\footnote{In fact, reality condition is sufficient to check that for $su(1,1)$, $$\eta={-2bt-a(-1+t^2+\beta^2)+c_3(1+t^2+\beta^2)\over 2\beta}={u \,k+v\, kt+w\, k(t^2+\beta^2)\over \beta},$$ under the mapping $z\rightarrow{-t+i(1-\beta)\over t+i(1+\beta)}$ with $c_1=a+ib$, $u={a+c_3\over 2 k}$, $v=-{b\over k}$ and $w={c_3-a\over 2k}$. Which exactly matches with \eqref{eq:generalG}.} that $c_1=c_2^*\in\mathbb{Z}$ and $c_3\in\mathbb{R}$. 
We now impose a boundary condition that the dilaton is minimised at the location $z = z_0$ i.e.
\begin{equation}
\left. \partial_z \eta(z, \bar{z}) \right|_{z = z_0} = \left. \partial_{\bar{z} } \eta(z, \bar{z}) \right|_{z = z_0} = 0   \label{minCond}
\end{equation}
which yields
\begin{eqnarray}
\eta(z, \bar{z}) & = & \frac{2 c_3 f (z - z_0)(\bar{z} - \bar{z}_0)}{(1 + f z_0 \bar{z}_0)(1 - f z \bar{z}) } + \eta_0   \nonumber \\
\eta_0 & = & \frac{c_3 (1 - f z_0 \bar{z}_0)   }{(1 + f \bar{z}_0 z_0)}    
\end{eqnarray}
This is precisely the spread complexity (\ref{spreadGen}) up to the overall factor $c_3$ and additive constant!  Indeed, this suggests that the dilaton solution may be identified with the spread complexity of the state $|z\rangle$.  What the condition (\ref{minCond}) is doing is identifying the reference state sourcing the Krylov basis.  The reference state is, of course, the point of minimal spread complexity and thus the minimisation condition.  With the identification of the classical dilaton background with the manifold of accessible target states, this is thus a natural boundary condition to impose on the dilaton.  \\ \\
To obtain a precise match between the dilaton and spread complexity we need to impose further boundary conditions.  The second condition, subtracting the additive constant $\eta_0$, is setting the spread complexity of the reference state to zero.  Put differently, the boundary condition (\ref{minCond}) matches the dilaton with a spread complexity cost function (\ref{spreadCost}) with the coefficients $c_n = m n + c$.  By subtracting the constant $\eta_0$ one fixes $c=0$ for the cost function coefficients.   
\\ \\
We need to set a final boundary condition to fix $c_3$, so that the shifted dilaton is matched with the $m=1$ spread complexity cost function.  We find that this boundary condition may be phrased in a physically interesting way.   In terms of the mass or energy of the gravitational system (\ref{JTEnergy}) we find that this third condition is equivalent to
\begin{equation}
\frac{1}{4}M R = -1
\end{equation}  
Recall that the scalar curvature of the FS manifold is inversely proportional to the weight of the highest weight state while the spread complexity is proportional to it.  Thus the total energy of the dilaton gravitational system must scale like the spread complexity with the weight of the highest weight state.  \\ \\
With these boundary conditions we are thus able to match the spread complexity of any target state (up to an additive constant) as the classical dilation of a JT gravity theory. We have already established that the Fubini-Study complexity geometry is obtained as the classical background geometry of this model. In this approach, the expectation value of symmetry generator $\expect{\hat{G}}$ on the quantum state manifold is the classical solution of dilaton in \eqref{classsoln}. For a detailed discussion interested reader can go to Appendix \ref{kill}. Taken together, these statements establish a neat link between spread complexity as a scalar field defined on the FS complexity geometry.   \\ \\
Note furthermore that our derivation are valid for any low-rank algebra of the form (\ref{algebra}) which encapsulates the cases of $su(1,1), su(2)$ and Heisenberg-Weyl.

\section{Conclusion and Discussion}\label{sec:conclusion}

The quantum computational complexity of a target and reference state pair can be defined in various different ways using various choices of cost function.  The effect of these cost functions is to add additional structure to the Hilbert space of accessible target states which provides a means to quantifying the difficulty of state preparation for each Hilbert space state. From a practical point of view, it is desirable to either find a physically ``most suited'' definition of complexity, or demonstrate that different definitions may be related in some way. In this paper we utilised a duality construction of quantum state manifolds to contribute towards this goal.  For low-rank algebras, we demonstrated that the Fubini-Study information metric and spread complexity of a target state can be obtained as the classical solutions of a single action.  Intriguingly, this action takes the form of a 2d dilaton gravity model.  Our results are in harmony with those that have already demonstrated a relation between spread complexity and the Fubini-Study metric volume \cite{Caputa:2021sib}.  Indeed, the identification made by those authors (relating the Krylov complexity operator to a FS metric isometry) can be understood in terms of a geometric formulation of quantum mechanics \cite{Ashtekar:1997ud} which underpins the duality construction.   In particular, it would be interesting to study the Jacobi group for which the spread complexity was studied quite recently \cite{Haque:2022ncl}.
 \\ \\
A natural question that arises is how the identification between the dilaton and spread complexity can and should be modified when higher dimensional symmetry groups are considered.  To this end we would like to make some additional comments.  Firstly, the Fubini-Study metric always encodes the group symmetries as metric isometries, but will in general not be maximally symmetric as is the case for the two-dimensional examples we considered.  Secondly, the Krylov complexity operator will, for higher dimensional cases, not be linear in the generators of the symmetry group.  We can expect, however, that the Krylov complexity operator may be expressible as an element of the enveloping algebra.  As such, it should still have an associated vector field on the manifold of quantum states related to its expectation value through (\ref{vecField}).  \\ \\
Taken together, these observations suggest that spread complexity may be identified as the classical solution of an appropriately defined scalar field action on the Fubini-Study background.  Our expectation, due to the points raised above, is that these actions may require the inclusion of a non-trivial energy momentum tensor.  Finding a general prescription to constrain these scalar field actions is an interesting (though likely also an involved) problem.  It would also be interesting to pursue the study the conjectured inequality of \cite{Avdoshkin:2022xuw} in this framework and pursue cases where the behavior of spread complexity differs from Nielsen complexity.  
 \\ \\ 
One should also note that, though the Fubini-Study metric for higher rank symmetry groups are not maximally symmetric in general, they may still permit maximally symmetric submanifolds.  There may be examples where the time-evolved states (and thus the Krylov basis) can be restricted to these submanifolds.  In these cases we expect the identification to simplify significantly and the appropriate action may even involve a vanishing energy-momentum tensor in these cases.  At this stage our expectation are only speculative, however, and we postpone a more careful analysis to future work.
\\ \\
We emphasise that the identifications made in this paper have all been on the classical solutions of the dilaton gravity action.  It would be fascinating to understand whether other quantities relevant to complexity may be packaged as the fluctuations of an appropriate action.  In particular, one may wonder whether the higher cumulants of the Krylov distribution (such as K-variance and K-skewness \cite{Parker:2018yvk, Barbon:2019wsy, Caputa:2021sib, Bhattacharjee:2022ave}) may be understood in this way.  On a related note, our focus in this paper has been on a particular choice of spread complexity (characterised by the increasing coefficients $c_n$) which we matched with a particular dilaton action, namely a linear dilaton action.  One may consider different models of dilaton gravity which may be related to different choices of these coefficients.   It is intriguing that a linear model of dilaton gravity yields classical solutions that may be matched with the spread complexity where the coefficients of the cost function also scale linearly. \\ \\
Finally, we remark that the connection between notions of quantum complexity and dilaton gravity, though intriguing, is not holographic in nature.  In particular, the quantum systems we have considered are not conformal field theories and the manifold of quantum states is of the same dimension as the dilaton gravity.  Nevertheless, it would be fascinating to study the duality construction for conformal field theories and whether notions of complexity for the CFT may be understood similarly. In this regard one can explore the known equivalence between JT gravity action and the Liouville gravity action containing both timelike and spacelike conformal fields 
 \cite{Suzuki:2021zbe}. Computation of Krylov complexity for this Liouville field theory from the two point correlation function will help us to verify the duality construction of dilaton from the field theoretic point of view. 

\section*{Acknowledgements}
The work of AC is supported by the European Union’s Horizon 2020 research and innovation programme under the Marie Sk\l{}odowska Curie grant agreement number 101034383. The work of AM is supported by the Ministry of Education, Science, and Technology (NRF- 2021R1A2C1006453) of the National Research Foundation of Korea (NRF). HJRvZ is supported by the “Quantum Technologies for Sustainable Devlopment'' grant from the National Institute for Theoretical and Computational Sciences (NITHECS). AC and HJRvZ acknowledges the 2021 postdoctoral retreat organised by University of the Witwatersrand with the support of National Research Foundation of South Africa.

\appendix

\section{Metric and the 2-form}\label{app:qstate}
We will pick up from the same setup as defined in \cref{sec:geoqu} for $\cH$ and $\cS$. In addition we will further define $\widetilde{\cS}$ as the set of rays $\{\widetilde{\ket{s}}=e^{i\theta}\ket{s}:\theta\in\mathbb{R}\}$ forming the projective Hilbert space $\mathcal{P}$. Here we will follow \cite{Provost:1980nc} and briefly review the basic ideas of quantum state manifolds, a first guess for metric can be defined through the square of the norm of two infinitesimally close vectors $\ket{s+ds}$ and $\ket{s}$ in $\cH$  as 
\begin{equation}\label{eq:normq}
	\norm{\Big(\ket{s+ds}-\ket{s}\Big)}^2=\braket{\Big(\ket{s+ds}-\ket{s}\Big)}{\Big(\ket{s+ds}-\ket{s}\Big)}.
\end{equation}
Upto first order in derivative expansion we have $\ket{\Big(\ket{s+ds}-\ket{s}\Big)}=\dow_i\ket{s}ds_i$, therefore one can write upto second order in derivative
\begin{equation}
	\norm{\Big(\ket{s+ds}-\ket{s}\Big)}^2=\bra{s}\overleftarrow{\dow_i}\overrightarrow{\dow_j}\ket{s}ds_ids_j,
\end{equation}
with $\dow_i\equiv{\dow\over \dow s_i}$. By separating the real and the imaginary part we will introduce the variables $\gamma_{ij}(s)$ and $\sigma_{ij}(s)$ as
\begin{equation}
	\bra{s}\overleftarrow{\dow_i}\overrightarrow{\dow_j}\ket{s}=\gamma_{ij}(s)+i\sigma_{ij}(s),
\end{equation}
which by definition of the hermitian products have the property
\begin{equation}
	\gamma_{ij}(s)=\gamma_{ji}(s)\quad \text{and}\quad \sigma_{ij}(s)=-\sigma_{ji}(s).
\end{equation}
This new parameters further simplifies the relation 
\begin{equation}
	\norm{\Big(\ket{s+ds}-\ket{s}\Big)}^2=\gamma_{ij}(s)\,ds_ids_j.
\end{equation}
Although, one can check that this tensor $\gamma_{ij}(s)$ as defined above has all the \emph{nice} properties for being a metric tensor but it fails to be one, if we want to build a manifold of physical states where both $\ket{s}$ and $\widetilde{\ket{s}}$ would define the same point on the manifold if
\begin{eqnarray}\label{eq:qrot}
	\widetilde{\ket{s}}=e^{i\theta(s)}\ket{s}.
\end{eqnarray}
Evidently, we have the relation 
\begin{equation}
\widetilde{\gamma_{ij}}(s)=\gamma_{ij}(s)+\beta_i\left(\dow_i \theta(s)\right)+\beta_j\left(\dow_i \theta(s)\right)+\left(\dow_i \theta(s)\right)\left(\dow_j \theta(s)\right),
\end{equation}
with
\begin{equation}
	\beta_j(s)=-i\bra{s}\overrightarrow{\dow_j}\ket{s}.
\end{equation}
As we are dealing with normalisable states here, we then have the relation
\begin{equation}
	\bra{s}\overrightarrow{\dow_i}\ket{s}=-\bra{s}\overleftarrow{\dow_i}\ket{s}.
\end{equation}
Adding with the relation $\bra{s}\overrightarrow{\dow_i}\ket{s}^\dagger=\bra{s}\overleftarrow{\dow_i}\ket{s}$, we have the property that $\beta_j(s)$ is real. Which leads one to the definition of a metric as 
\begin{equation}
	g_{ij}(s)=\gamma_{ij}(s)-\beta_i(s)\beta_j(s),
\end{equation}
which by construction is invariant under \eqref{eq:qrot}. Following \cite{Provost:1980nc}, one can also show that locally one can define the 2-form $\sigma$ and 1-form $\beta$ as
\begin{equation}
	\sigma=\sigma_{ij}(s) ds_i\wedge ds_j,\quad \text{and}\quad \beta=\beta_i(s)ds_i\quad\text{with}\quad \sigma=d\beta.
\end{equation}
Furthermore by resorting to the following relation, upto second order in derivative
\begin{equation}
	\left[\braket{\Big(\ket{s+ds}-\ket{s}\Big)}{\Big(\ket{s+ds}-\ket{s}\Big)}\right]-\left|\bra{s}{\Big(\ket{s+ds}-\ket{s}\Big)}\right|^2=g_{ij}(s)ds_ids_j,
\end{equation}
it is evident that
\begin{equation}
	dl^2=g_{ij}ds_ids_j,
\end{equation}
is always positive definite by Schwartz's inequality. One can then write the following relations for $\gamma_{ij}$, $\sigma_{ij}$ and $\beta_i$ in a simpler form as
\begin{equation}\label{eq:defapp}
	\beta_j(s)\equiv -i \dow_j'\braket{s}{s'}|_{s=s'}\quad \text{and}\quad \gamma_{ij}(s)+i\sigma_{ij}(s)\equiv \dow_i\dow_j'\braket{s}{s'}|_{s=s'}
\end{equation}
which further implies \eqref{eq:sympdef}. Therefore, by knowing the inner product between any two elements of $\widetilde{\cS}$ the metric $g$ and the 2-form $\sigma$. At this point one of the interesting questions to ask is the connection between the phase space of a classical system (say $\Gamma$) and the quantum state space manifold. Both $\Gamma$ and $\mathcal{P}$ have a symplectic structure but the quantum state space manifold also have a Riemannian metric due to the K\"ahler structure. In general even if the classical phase space is finite dimensional the \emph{quantum phase space} or $\mathcal{P}$ is still infinite dimensional. Interestingly, one can show that the quantum phase space is basically a trivial bundle over the classical phase space \cite{Ashtekar:1997ud}. Therefore, corresponding to each choice of a state (or a point $p\in\mathcal{P}$) there is a cross section equivalent to the classical phase space $\Gamma$. Moreover, it turns out that the quantum states that lies within this cross sections are precisely the generalised coherent states\cite{Perelomov}.

To illustrate the above points one can take for example the Glauber coherent  states for a harmonic oscillator as 
\begin{equation}
	\ket{\alpha}=e^{-{1\over 2}|\alpha|^2}\sum_{n=0}^\infty {\alpha^n\over (n!)^{1/2}}\ket{n}
\end{equation}
with $\alpha\in\mathbb{C}$. The inner product in this case is
\begin{equation}
	\braket{\alpha}{\alpha'}=\expa{\bar{\alpha}\alpha'-{1\over 2}|\alpha|^2-{1\over 2}|\alpha'|^2}.
\end{equation}
Hence using \eqref{eq:sympdef} or \eqref{eq:defapp} we can write the relation
\begin{equation}
	\begin{split}
	dl^2&=d\alpha_1^2+d\alpha_2^2=d\rho^2+\rho^2d\phi^2,\\
	\sigma&=2\,d\alpha_1\wedge d\alpha_2=2 \rho d\rho\wedge\d\phi,
	\end{split}
\end{equation}
where we have parametrised $\alpha$ as $\alpha=\alpha_1+i\alpha_2=\rho e^{i\phi}$. Therefore the manifold corresponding to Glauber coherent state is a 2-dimensional plane.

\subsection{Isometries and dynamical symmetries}
The definition for the metric and the 2-form in \eqref{eq:defapp} depends on the inner product on $\cH$, therefore they are also expected to be invariant under unitary transformations that keeps $\widetilde{\cS}$ invariant. This kind of transformations can be written as
\begin{eqnarray}\label{eq:transU}
	\hat{U}\ket{s}=e^{i\phi(s)}\ket{u(s)}\quad \text{where}\quad s\rightarrow u(s),
\end{eqnarray}
and $\phi(s)$ is some scalar function. From the change of inner product under this mapping we then have the relation
\begin{equation}
	g_{ij}(s)={\dow u_k\over \dow s_i}{\dow u_l\over \dow s_j}g_{kl}(u)
\end{equation}
implying that \eqref{eq:transU} defines a isometry of the metric. Following \cite{Kriel:2015tga}, one can call them \emph{dynamical symmetries}. This rather simple exercise reveals the dual role of the inner product $\braket{s}{s'}$. In the algebraic side, the dynamical symmetries implies specific transformation properties of $\braket{s}{s'}$, restricting their functional form. In the geometric side, $\braket{s}{s'}$ acts as the potential of the K\"ahler manifold from which the metric and the 2-form is derived. Here the dynamical symmetries defines the isometries of the manifold. Hence, one can realise that $\braket{s}{s'}$ acts as the \emph{missing link} between the quantum mechanical symmetries of a system and the isometries defining the geometry of the system. 

\section{Manifolds for non-normalisable states}\label{UnnormalisableRefStates}

So far, we have discussed normalisable and non-normalisable states more or less interchangeably. Although, the notion of symmetry transformation and the vector field properties as discussed before remains identical, the caveat comes through  \eqref{eq:sympdef} for the definition of metric and the symplectic structure due the singular nature of the inner product. This problem can be circumvented by realising that in fact, the set of potentially non-normalisable vectors $\cS_\infty=\{|s)\}$, can have a enhanced set of symmetry transformations.\\ \\
To understand the above point, one can consider a one dimensional manifold corresponding to the trajectory of a state through the Hilbert space under the time evolution generated by the Hamiltonian. From the discussions above and \cref{app:qstate}, one can state that the one-dimensional manifold generated through this evolution will only have a single isometry direction due to the fact that $\cS$ only have a single continuous symmetry under time translation. The weakness of not having a well defined metric become the saviour for non-normalisable states in this situation. For non-normalisable states one can then look for a possibly larger set of dynamical symmetries of the system which can be carry forwarded as the symmetries of $\cS_{\infty}$. To this vein one can start with a non-normalisable state of a one-dimensional system as the reference state and generate a manifold of \emph{normalisable} states where the enhanced symmetries can act as isometries. \\ \\
The simplest way to achieve the above is to introduce a regularisation coordinate $\beta$ following \cite{Kriel:2015tga}, which can be introduced as the imaginary part of \emph{time} such that we have a family of states
\begin{equation}
	|t,\beta)=e^{it\hat{H}}\ket{\beta}=e^{i(t+i\beta)\hat{H}}|\phi_0);\quad (t,\beta)\in \mathbb{R}\times (\beta_0,\infty)
\end{equation}
with $|\phi_0)$ being our reference state. Here we are also assuming that $e^{-\beta \hat{H}}$ is the precise factor needed to render $|t,\beta)$ normalisable as $\beta>\beta_0$. Without loss of generality one can now set $\beta_0=0$ and think of the state manifold being divided in two parts, the infinite norm ``boundary'' states at $\beta=0$ and the finite norm states inside the ``bulk'' $\beta>0$. The regularisation coordinate $\beta$ appears as some energy scale in the system and \cite{Kriel:2015tga} showed that $\beta$ actually plays the analogous role of radial bulk coordinate as in AdS/CFT.\\ \\
One can now use \eqref{eq:sympdef} and calculate the bulk metric and the 2-form as 
\begin{equation}\label{eq:metric2d}
	g=C(\beta)\left(dt^2+d\beta^2\right)\quad \text{and}\quad \sigma=2 C(\beta)dt\wedge d\beta,
\end{equation}
where 
\begin{equation}
	C(\beta)=\big<\beta|(\delta\hat{H})^2|\beta\big>={1\over 4}F''(\beta),\quad F(\beta)=\log Z(\beta)\quad \text{and}\quad Z(\beta)=(t,\beta|t,\beta).
\end{equation}
The scalar curvature of this manifold can be calculated to be 
\begin{equation}
	\cR=-{\dow_\beta^2\loga{C(\beta)}\over C(\beta)}.
\end{equation}
The metric $g$ here is also a Fubini-Study metric since declaring $\tau=t+i\beta$ to be the complex coordinate one can see that both $g$ and $\sigma$ can be calculated using \eqref{eq:cmplxdef} through the K\"ahler potential ${1\over 2}\log(\tau|\tau)$ with $|\tau)\equiv|t,\beta)$. By definition $Z(\beta)$ will diverge at the ``boundary'' $\beta=0$. Assuming $Z(\beta)=\beta^{-p}\expa{f(\beta)}$,  with $p>0$ and $f(\beta)$ being analytic at $\beta=0$, we have the relation
\begin{equation}
	C(\beta)={p\over 4 \beta^2}+{f''(0)\over 4}+\cO(\beta)\quad \text{and}\quad \cR=-{8\over p}-{16 f'''(0)\over p^2}\beta^3+\cO(\beta^4).
\end{equation}
Therefore the conformal factor of the metric diverges as $\sim \beta^{-2}$ whereas the scalar curvature approaches constant negative value which is the telltale sign of asymptotically Euclidean AdS space. Since $Z(\beta)$ encapsulates symmetries of both the Hamiltonian and the reference state $|\phi_0)$, one can easily show that for a situation where $\hat{H}$ and $|\phi_0)$ are both invariant under a scale transformation,  the metric turns out to be exactly $AdS_2$.

\subsection{Killing Vectors}\label{kill}
In this part we will consider systems governed by the $su(1,1)$ algebra for simplicity but one can extend this discussion without much modifications for $su(2)$ and Heisenberg-Weyl and other low rank algebras as well. We consider the generators of $su(1,1)$ being $\hat{K}_0$, $\hat{K}_+$ and $\hat{K}_-$ with the Casimir operator $\hat{C}=k(k-1)\hat{I}$, with $k$ labelling the particular irreducible representations under consideration. In this case, the Hilbert space is spanned by the states $\{|k,n\big>\}$ ($n\in[0,\infty])$, where the action of the generators can be defined as 
\begin{equation}
	\begin{split}
	\hat{K}_0|k,n\big>&=(k+n)\ket{k,n}\\
	\hat{K}_+\ket{k,n}&=\sqrt{(n+1)(2k+n)}\ket{k,n+1}\\
	\hat{K}_-\ket{k,n}&=\sqrt{n(2k+n-1)}\ket{k,n-1}.
	\end{split}
\end{equation}
This generators can be thought of as the three conformal generator of conformal quantum mechanics \cite{deAlfaro:1976vlx} as 
\begin{equation}
	\begin{split}
	\hat{K}&=\hat{K}_0+{1\over 2}\left(\hat{K}_++\hat{K}_-\right)\\
	\hat{T}&=\hat{K}_0-{1\over 2}\left(\hat{K}_++\hat{K}_-\right)\\
	\hat{D}&={i\over 2}(\hat{K}_+-\hat{K}_-).
	\end{split}
\end{equation}
 One can now consider $\hat{T}$ as the Hamiltonian and consider the time evolution with the states $\ket{t}=e^{it\hat{T}}\ket{\phi_0}$. To endow the reference state with the full set of symmetries of $su(1,1)$ we will consider a non-normalisable family of states defined as 
\begin{equation}
	|\phi_0)=e^{-\hat{K}_+}\ket{k,0}
\end{equation}
which has infinite norm and is a simultaneous eigenstates of $\hat{K}$ and $\hat{D}$ \cite{Kriel:2015tga}. Interestingly, these are the precise family of states $\{|t)=e^{it\hat{T}}|\phi_0)\}$ considered in \cite{Chamon:2011xk}, in connection with the $AdS_2/CFT_1$ correspondence. It is shown to be \cite{Chamon:2011xk,Kriel:2015tga}
\begin{equation}
	(t'|t)=\left[{i\over 2(t-t')}\right]^{2k}.
\end{equation}
which basically dictates a two point function for a field with conformal dimension $k$. As expected, this inner product diverges as $t\rightarrow t'$, therefore to get a valid geometry we have to introduce the regularisation coordinate $\beta$ as $t\rightarrow \tau=t+i\beta$ and study the sates $|\tau)\equiv|t,\beta)=e^{i\tau \hat{T}}|\phi_0)$. With this regularisation we have $(\tau|\tau)=(4\beta)^{-2k}$ which is finite for $\beta>0$. Using \eqref{eq:metric2d}, the resulting metric for this two dimensional manifold spanned by $t$ and $\beta$ turned out to be the $AdS_2$ metric
\begin{equation}\label{eq:qsmetric}
	ds^2={k\over 2\beta^2}\left[dt^2+d\beta^2\right]
\end{equation}
with the scalar curvature $\cR=-{4\over k}$. The three isometries of these metric are the three symmetries generated through the three conformal transformation of $su(1,1)$. Although in this discussion we have taken $\hat{T}$ to be the Hamiltonian, but as argued in \cite{Kriel:2015tga} the geometry is determined by the dynamical symmetry group and not on the particular element of the algebra chosen as the Hamiltonian. Therefore the above observation is applicable to any $su(1,1)$ Hamiltonian.

By identifying $\hat{V}_1=\hat{K}$, $\hat{V}_{-1}=\hat{T}$ and $\hat{V}_0=\hat{D}$, one can verify that $\{i\hat{V}_n\}$ satisfy the global part of the Virasoro algebra as
\begin{equation}\label{eq:virasoroglobal}
	\left[i\hat{V}_n,i\hat{V}_m\right]=(n-m)i\hat{V}_{n+m}.
\end{equation}
Hence in principle, one can utilise this same machinery to study conformal field theory states as well. As discussed in \eqref{KillingExpectationVal}, the Killing vector fields $\{X_{\hat{V}_\pm},X_{\hat{V}_0}\}$ would provide a representation of $su(1,1)$ acting on the the scalar fields defined on $\cS$. We already have the result from \eqref{KillingExpectationVal} that for an arbitrary generator $\hat{G}$ the associated vector field is $X_{\hat{G}}=k^\tau\dow_\tau+k^{\bar{\tau}}\dow_{\bar{\tau}}$ where $k^\tau=-{1\over 2}\sigma^{\tau\bar{\tau}}\dow_{\bar{\tau}}\expect{\hat{G}}$. For this generator to be a conformal transformation, $k^\tau$ must satisfy $\dow_{\bar{\tau}}k^\tau=0$ to be holomorphic. This condition is equivalent to claiming $k_a$'s are solution of the conformal Killing equation
\begin{equation}
	\nabla_{a}k_{b}+\nabla_bk_a=(\nabla\cdot k)g_{ab}.
\end{equation}
Therefore we conclude that in general the solution for the scalar field is a linear combination of $\{\expect{\hat{V}_n}:n\in(0,1,-1)\}$ plus an arbitrary holomorphic function\footnote{The general \emph{complex} solution of the conformal Killing equation in $\{\expect{\hat{V}_n}:n\in \mathbb{Z}\}$, but for real scalar fields one have to restrict $n\in(-1,0,1)$ \cite{Kriel:2015tga}. }. One can now obtain \cite{vanZyl:2015dsa} for $\beta>0$
\begin{equation}
	\begin{split}
	\bra{\beta,t}\hat{V}_{-1}\ket{\beta,t}&={k\over \beta}\\
	\bra{\beta,t}\hat{V}_{0}\ket{\beta,t}&={kt\over \beta}\\
	\bra{\beta,t}\hat{V}_{1}\ket{\beta,t}&={k(t^2+\beta^2)\over \beta}.
	\end{split}
\end{equation}
Therefore for an arbitrary $su(1,1)$ symmetry generator $\hat{G}=u\hat{V}_{-1}+v\hat{V}_0+w\hat{V}_{1}$ the associated scalar field is 
\begin{equation}\label{eq:generalG}
	\expect{\hat{G}}={u\, k+v\,kt+w\,k(t^2+\beta^2)\over \beta}.
\end{equation}

\bibliographystyle{jhep}
\bibliography{bib_complexity}{}

\end{document}